# Intelligent Anti-Money Laundering Solution Based upon Novel Community Detection in Massive Transaction Networks on Spark

Xurui Li[1,2], Xiang Cao[2], Xuetao Qiu[1], Jintao Zhao[1], Jianbin Zheng[1]
*China UnionPay[1], Shanghai, China*
*School of Computer Science, Fudan University[2], Shanghai, China*
*Corresponding author: xurui.lee@msn.com; lixurui@unionpay.com*

*Abstract*—Criminals are using every means available to launder the profits from their illegal activities into ostensibly "legitimate" assets. Meanwhile, most commercial anti-money laundering systems are still rule-based, which cannot adapt to the ever-changing tricks. Although some machine learning methods have been proposed, they are mainly focused on the perspective of abnormal behavior for single accounts. Considering money laundering activities are often involved in gang criminals, these methods are still not intelligent enough to crack down on criminal gangs all-sidedly. In this paper, a systematic solution is presented to find suspicious money laundering gangs. A temporal-directed Louvain algorithm has been proposed to detect communities according to relevant anti-money laundering patterns. All processes are implemented and optimized on Spark platform. This solution can greatly improve the efficiency of anti-money laundering work for financial regulation agencies.

## I. INTRODUCTION

Money laundering (ML) refers to the use of a series of financial proceeds to cover up the illegal source of funds from corruption, fraud, smuggling and other forms of crime, making the money appear legitimate [1]. With the increasing rampant of upstream crime, ML is posing a more serious threat to financial institutions as well as national security. How to effectively detect abnormal financial activities has become a huge challenge faced by governments and financial institutions. Anti-money laundering (AML) systems have been deployed by some governmental and financial institutions to combat with criminals. Nevertheless, most of the AML systems are still rule-based, suffering from numerous of drawbacks such as insufficient data processing capability, lack of pattern recognition function and easy to be avoided [2]. Some machine learning technologies such as classification and sequence analysis based on historical transaction information for special account have been carried out, improving the efficiency of AML work to some extent [3]-[5]. However, analysis on isolated accounts may still lose important related information, because ML activities are often involved in gang crimes. AML systems should be more intelligent to be able to detect the suspicious ML gangs quickly and accurately.

Graph mining methods are often used to explore the associations between individuals. Furthermore, community detection algorithm can be an effective method to find ML gangs [6]. While considering the huge volume of transactions, elements with little suspicion should be filtered out before community detection. On the other hand, transfer time and direction are two crucial factors during anti-money laundering processes. Further complex ML crimes can be predicted and prevented if inherent evolution law of the transaction structure is grasped in early times. Nevertheless, current existing methods such as GN (Girvan-Newman) [7], CPM (Clique Percolation Method) [8] or Louvain [9] algorithms can't handle the temporal-directed network well.

In this paper, a comprehensive method for detecting suspicious ML gangs in massive transaction networks has been presented. Noise information is filtered out first to reduce the total computation cost. An algorithm incorporated with rich AML experience has been proposed to detect communities. The algorithm has also been parallelized and optimized in Spark GraphX platform, which was applied to deal with the massive real transaction data. At last, the most suspicious ML communities can be picked out by reordering the calculated risk score. This solution has been proved to be a powerful auxiliary tool for monitoring department carrying out anti-money laundering work.

## II. DESIGN AND IMPLEMENTATION

### A. Select effective maximal connected subgraph in massive transaction networks

For a large financial institution like UnionPay, there are tens of millions of transactions every day. It is hard and a waste of time to divide all transactions into communities. Meanwhile, criminals always want to launder their money in a relatively short period of time. Thus, we focus on the transaction data in a certain time period $P_T$. The transfer net can be established according to the real transaction data in $P_T$. Each account which has transaction record during $P_T$ corresponds to a node, and the transfers between these accounts are treated as edges. First we need to merge the edges between node pairs with the same source and destination node. This can be easily done using the *groupEdges* function in Spark GraphX. The sum expression was used to measure money and times properties, while the average expression was used to measure the transfer time point property. Note that the edges should be repartitioned before using *groupEdges* function in Spark.

GraphX exposes a triplet view, which logically joins the vertex and edge properties yielding an *EdgeTriplet* RDD. Supposing there are n edges in the whole graph. To express the key idea of our algorithm to the point, we only use crucial factors such as transfer money and times to calculate the primitive edge weight here. For a certain *EdgeTriplet* i, the total transfer money and times is defined as $M_i$ and $C_i$, respectively. As we know, the standardization of a variable $A_i$ is expressed as: $\bar{\bar{A}}_i = (A_i - \bar{A}_i)/\sigma_A$, where $\bar{A}_i =$

$(\sum_{i=1}^{n} A_i)/n$ and $\sigma_A = \sqrt{(\sum_{i=1}^{n}(A_i - \overline{A}_i)^2)/n}$. The primitive edge weight $W_{Bi}$ then can be described as: $w_{Bi} = e^{\omega_m \cdot \overline{M}_i + \omega_c \cdot \overline{C}_i}$, where the symbols contain ω represent the weight distribution ratio for each variables respectively, with their sum at 1. These ratios can be adjusted according to business requirements. For demonstration, these ratios are averaged allocated. The ratios expressed by symbol ω in the following context are disposed in the same way. Note that $W_{Bi}$ increases with $M_i$ and $C_i$,

After merging edges, it can be found that there are many isolated edges, whose source node and destination node only connect to one edge. These edges have a great disruption to the following ML works, and can be filtered out by using *subgraph* function. Then the *ConnectedComponents* function in Spark can be applied to divide the graph into different maximal connected subgraph (MCS) [10]. According to domain knowledge, criminals are always trying to make the transfer network complicated to conceal the true origin and ownership of the proceeds of their criminal activity. A MCS with very little scale is unlikely to be a ML gangs. However, a MCS may be associated with some large merchants and has little possibility to be a ML gangs if it is particularly large and complex. Thus, the MCSs can be filtered using the following formula: $V_{\theta 1} < V_{mcs} < V_{\theta 2}$, where $V_{mcs}$ is the threshold scale of MCSs. In addition, the node whose degree exceeds a threshold $D_\theta$ is defined as "hub node". A MCS with higher ML possibility should have more hub nodes. MCSs can be further filtered by $N_{hubs} > N_\theta$, where $N_{hubs}$ is the number of hub nodes in a MCS and $N_\theta$ is the threshold.

## B. Community detection according to ML characteristics

Analyzing a MCS directly may be complex and less effective because core ML structures are always mixed with some normal transactions. Thus further divisions of the MCSs were made here by using community detection algorithm. Louvain algorithm with both relatively good speed and performance was selected to implement the community detection [11]. However, the original Louvain algorithm is a common method, which may not have much effect in the field of ML. So a temporal-directed Louvain (TD Louvain) algorithm has been proposed here to detect the communities for AML goals. The details of the algorithm are described as below.

a) Edge weight optimization by node correction

The original modularity-based Louvain algorithm mainly measures the impact of edge weight to community. However, it overlooks the weight of node, which is very important in ML networks. Learning from the idea of PageRank [12], assuming node A is already known as an important node, all edges directly connected with A should be relative suspicious, no matter how little the transfer money and times are. For example, a small transaction between A and B is perhaps to be a pre-tentative transaction, if not found, a sequence of large transactions may be followed. For an *EdgeTriplet* i with starting node src and terminating node dst, the node correction for src is $\sigma_s = e^{\omega_{Mv} \cdot \overline{\overline{M}}_s + \omega_{Cv} \cdot \overline{\overline{C}}_s + \omega_{Dv} \cdot \overline{\overline{D}}_s}$. Here $\overline{\overline{D}}_s$, $\overline{\overline{M}}_s$ and $\overline{\overline{C}}_s$ are standardized degree, money and times for node src. The correction for node dst $\sigma_d$ is calculated in the same way. Thus, current edge weight can be expressed as: $w_{Ni} = \sigma_s * \sigma_d * w_{Bi}$.

b) Temporal correction for edge weight

Criminals are always trying to centralize or decentralize their illicit money in a short period. To deal with this problem, a graph-based pattern matching method is introduced here [13]. The average time point of all inbound and outbound transfers has been calculated as $\overline{t_A^{in}}$ and $\overline{t_A^{out}}$ respectively. For a directed *EdgeTriplet* i with a starting node src and a terminating node dst, the average transfer time point property for edge src → dst has already been calculated using *groupEdges* function as $T_{s \to d}$. If the in-degree of node src is larger than out-degree ($\text{Deg}_s^{in} > \text{Deg}_s^{out}$), the edge src → dst is likely to follow a pattern $P_1$ called *"centralized out after multi transfer in"*, whose weight should be modified. Otherwise, no additional correction will be made. Focusing on the word "after", it means an edge completely follows pattern $P_1$ if $T_{s \to d} > \overline{t_s^{in}}$, with its edge weight promoted. Yet, if $T_{s \to d} < \overline{t_s^{in}}$, it can be inferred that the risk of this edge may be very low, whose weight should be reduced.

A correction factor for node src is defined as: $\theta_s = e^{\beta_s \cdot \tau_s^{out}}$. Here $\beta_s = (\text{Deg}_s^{in} - \text{Deg}_s^{out})/\text{Deg}_s$ and $\tau_s^{out} = P_T/(T_{s \to d} - \overline{t_s^{in}})$. The denominator $\text{Deg}_s = (\text{Deg}_s^{in} + \text{Deg}_s^{out})$ appears in the expression of $\beta_s$ is for the purpose of numerical standardization, preventing the weight correction factor from growing too large. The $\tau_s^{out}$ being divided by the whole time interval $P_T$ not only aims at standardization, but also ensures that the absolute value of correction coefficient is larger if $T_{s \to d}$ is closer to $\overline{t_s^{in}}$ (because this condition is more suspicious). As can be seen, if the edge src → dst satisfies $\beta_s > 0$, $\text{Deg}_s^{in} > \text{Deg}_s^{out}$ and $T_{s \to d} > \overline{t_s^{in}}$ simultaneously, correction factor $\theta_s > 1$. For other cases, $\theta_s \leq 1$. In a similar way, if the destination node dst follows a pattern $P_2$ called *"centralized in edge before multi transfer out"* pattern, the weight of edge src → dst should also be enhanced. Another weight correction factor for terminating node can be defined as: $\theta_d = e^{\beta_d \cdot \tau_d^{in}}$ in the same way, where $\beta_d = (\text{Deg}_d^{in} - \text{Deg}_d^{out})/\text{Deg}_d$ and $\tau_d^{in} = P_T/(T_{s \to d} - \overline{t_d^{out}})$. Then the final edge weight of src → dst in *EdgeTriplet* i can be expressed as:

$$W_{Ei} = \begin{cases} \theta_s \cdot \theta_d \cdot w_{Ni}, & \beta_s > 0 \text{ and } \beta_d < 0 \\ \theta_s \cdot w_{Ni}, & \beta_s > 0 \text{ and } \beta_d > 0 \\ \theta_d \cdot w_{Ni}, & \beta_s < 0 \text{ and } \beta_d < 0 \\ w_{Ni}, & \beta_s < 0 \text{ and } \beta_d > 0 \end{cases} \quad (1)$$

An instance of the above idea is shown in Fig. 1. The source node A of edge $E_2$ satisfies $\beta_A > 0$ and the destination node B satisfies $\beta_B < 0$, while other edges do not meet the two conditions simultaneously. So edge $E_2$ should have the most weight corrections among all edges. If node A satisfies pattern $P_1$ and B satisfies pattern $P_2$ after taking timing factor into account, then the weight of edge $E_2$ obtains the most reinforcement from corrections if other conditions are the same for all edges. Thus the edge weights in the following selections all use $W_{Ei}$ calculated by Eq (1).

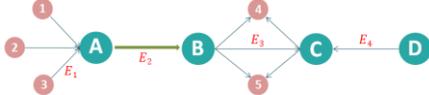

*Fig. 1. An instance to clarify temporal correction*

c) Directed optimization for modularity

Louvain method implements community detection in a network by maximizing modularity [14]. However, the asymmetry of information caused by the direction of edges has not been taken into account in original algorithms. In directed graph theory, it is considered that if node i has "low in-degree and high out-degree" while node j is just opposite, then link $j \to i$ is more abnormal than $i \to j$ [15]. In another word, link $j \to i$ will play a more significant role for community detection. This point also has a practical meaning for AML. Focusing on the fund flow illustrated in Fig. 2, there is no reason to doubt that edge $j \to i$ is more suspicious than $i \to j$, if all other conditions are the same. It is because a more structurized transactions can be formed by edge $j \to i$ along with other related edges. Here, the edge $j \to i$ is very likely to be an intermediary channel between a decentralized in and decentralized out transfers.

Following the above conception, it has previously been thought that the expression of $k_i k_j$ during modularity calculation can be modified to $k_i^{in} k_j^{out}$ [15]. However, this modification is not so accurate. Consider a source node i with no edges connected in, then the $k_i^{in} = 0$ suggests that no matter how large $k_j^{out}$ is, the effect of edge $i \to j$ does not change. The same situation can happen for a destination node j with no edges out. Accordingly, a proportional power function revised factor for node n has been defined as: $\delta_n = (k_n^{in} - k_n^{out})/k_n$, where $k_n^{in}$, $k_n^{out}$ and $k_n$ is the weight sum of edges linked into, linked out and linked with node n, respectively. Then the $k_i k_j$ expression can be revised into $e^{\delta_i - \delta_j} k_i k_j$, and the modularity can be expressed as:

$$Q_D = \frac{1}{2m}\sum_{i,j}\left[A_{ij} - \frac{e^{\delta_i - \delta_j} k_i k_j}{2m}\right]\delta(c_i, c_j) = \frac{1}{2m}\left[\sum_{i,j} A_{ij} - \frac{\sum_i e^{\delta_i} k_i \sum_j e^{-\delta_j} k_j}{2m}\right]\delta(c_i, c_j) = \sum_c\left[\frac{\sum W_{ec}}{2m} - \left(\frac{1}{2m}\right)^2 \sum M_c\right] \quad (2)$$

Here $A_{ij}$ represents the weight of the edge between i and j. $m = \frac{1}{2}\sum_{i,j} A_{ij}$ is the sum of edge weights in the whole graph. $c_i$ is the community where node i belongs to, and $k_i$ is the sum of all edge weights attached to node i. If $c_i = c_j$, $\delta(c_i, c_j)$ is equal to 1, otherwise the function value is 0. The corresponding matrix for each community $M_c$ is:

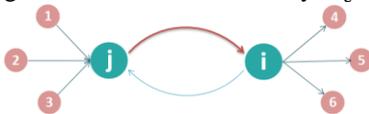

*Fig. 2. A fund flow illustration of directed optimization*

$$M_c = \begin{bmatrix} e^{\delta_1 - \delta_1} k_1 k_1, & e^{\delta_1 - \delta_2} k_1 k_2, & \cdots & e^{\delta_1 - \delta_N} k_1 k_N \\ e^{\delta_2 - \delta_1} k_2 k_1, & e^{\delta_2 - \delta_2} k_2 k_2, & \cdots & e^{\delta_2 - \delta_N} k_2 k_N \\ \vdots & \vdots & \ddots & \vdots \\ e^{\delta_N - \delta_1} k_N k_1, & e^{\delta_N - \delta_2} k_N k_2, & \cdots & e^{\delta_N - \delta_N} k_N k_N \end{bmatrix} \quad (3)$$

$\sum M_c$ represents the sum of all elements in matrix $M_c$. $\sum_c$ means accumulating in the original community c. It can be found that if $k_i^{in}$ or $k_j^{out}$ is very small, then the contribution to modularity of edge $j \to i$ can be greater than that of edge $i \to j$.

When the revised modularity has been defined, the iterative algorithm can be carried out to maximize the modularity. Detail steps are as follows:
1) Initialize the community tag for each node by using its own node tag.
2) Traversing all nodes by attempting to allocate each node i to the community where its neighbor node resides. Calculate the difference $\Delta Q_D$ before and after each allocation and pick out the allocation with maximum $\Delta Q_D$. If maximum $\Delta Q_D$ is positive, then actualize this allocation; Otherwise, no change will be done. The formula for $\Delta Q_D$ is revised as:

$$\Delta Q_D = \left[\frac{\sum W_{ec} + k_i^c}{2m} - \frac{1}{(2m)^2}\sum M_{cnew}\right] - \left[\frac{\sum W_{ec}}{2m} - \frac{1}{(2m)^2}\sum M_c - \frac{e^{\delta_i - \delta_i} k_i k_i}{(2m)^2}\right] = \frac{k_i^c}{2m} - \frac{\sum M_{cnew} - \sum M_c - k_i k_i}{(2m)^2} = \frac{k_i^c}{2m} - \frac{\Theta_i}{(2m)^2} \quad (4)$$

Where $k_i^c$ is the total weight of edges formed between node i and all nodes in community c; $\Theta_i = k_i e^{\delta_i}\sum_c k_j e^{-\delta_j} + k_i e^{-\delta_i}\sum_c k_j e^{\delta_j}$, which corresponding to the sum of all colored elements in $M_{cnew}$. The detail expression of $M_{cnew}$ is: $M_{cnew} =$

$$\begin{bmatrix} e^{\delta_1 - \delta_1} k_1 k_1, & e^{\delta_1 - \delta_2} k_1 k_2, & \cdots & e^{\delta_1 - \delta_N} k_1 k_N, & e^{\delta_1 - \delta_i} k_1 k_i \\ e^{\delta_2 - \delta_1} k_2 k_1, & e^{\delta_2 - \delta_2} k_2 k_2, & \cdots & e^{\delta_2 - \delta_N} k_2 k_N, & e^{\delta_2 - \delta_i} k_2 k_i \\ \vdots & \vdots & \ddots & \vdots & \vdots \\ e^{\delta_N - \delta_1} k_N k_1, & e^{\delta_N - \delta_2} k_N k_2, & \ddots & e^{\delta_N - \delta_N} k_N k_N, & e^{\delta_N - \delta_i} k_N k_i \\ e^{\delta_i - \delta_1} k_i k_1, & e^{\delta_i - \delta_2} k_i k_2, & \cdots & e^{\delta_i - \delta_N} k_i k_N, & e^{\delta_i - \delta_i} k_i k_i \end{bmatrix} \quad (5)$$

3) Repeat the step 2) until the community tag of all nodes does not change.
4) Compress nodes with same community label into a new node. The total edge weight among inner nodes in original community is transformed into self-link weight of new node; The weight between communities is transformed into that between new nodes.
5) Repeat the step 2) until the modularity of the whole graph does not change.

*C. Algorithm parallelization based on Spark GraphX*

The original Louvain algorithm is not suitable for implementation on a distributed platform like Spark directly. Therefore, the algorithm needs to be optimized in parallel. The main idea of parallelization is to update the information of multiple nodes synchronously according to that of neighbor nodes in last iteration [16]. Details are as follows:

The 5 steps of the Louvain algorithm described above can be divided into two stages here: the original steps 1 to 3 are assigned to the first stage, which is to set the community tag of each node until no change is made; Steps 4 and 5 are assigned to the second stage, which is to build a new graph and re-perform the first stage until the whole modularity no longer increases. Parallelization can be implemented on each stage, respectively.

a) Parallelization for the 1st stage

A data structure has been defined as $Info_{(i,j)} = \{k_i, k_i^{in}, k_i^{out}, c_i, k_j, k_j^{in}, k_j^{out}, c_j\}$ to record the relevant information for each node pair (i,j) on each iteration. Here $k_i$ is the total weight of edges connected with node i, $k_i^{in}$ is total weight of edges linked into node i, $k_i^{out}$ is total weight of edges linked out from node i, and $c_i$ is the community tag for node i. The same is for that of node j.

The parallelization can be achieved by using the *aggregateMessages* function in newest Spark, instead of *mapReduceTriplet* in old versions. $Info_{(i,j)}$ instances for each node and all its neighbor nodes are generated during map phases, while the neighbor information is assembled into an array for each node in reduce phases. The new community tag for each node can be determined when all its neighbor information is obtained.

b) Extra corrections for 1st stage parallelization

In above process, some problem may be encountered during the parallelization of 1st stage. As shown in 1st part of Fig. 3, the original nodes a, b, c, d, i, j belong to their own communities. In a round of iterations, there is a certain risk that node i may be assigned to the community where node j originally located, meanwhile node j is assigned to the community where node i originally located. Thus a "Community Swap" problem arises, as shown in the 2nd of part of Fig. 3. On the other hand, the 3rd part of Fig. 3 shows another potential issue called "Ascription Lag". In this situation, nodes a and b are assigned to the community where node j originally located, while node j itself alters its community ascription to i. This is apparently unreasonable, and may even generate communities with isolated node.

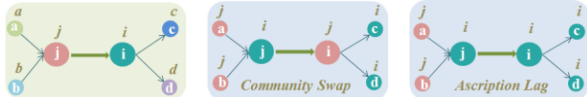

Fig. 3. "Community Swap" and "Ascription Lag" issues during parallelization processes

What we have done here is to make extra judgments after the 1st stage on each iteration. Nodes with community tag changed will be focused on. If there is a "Community Swap" in current round, then the two concerned nodes will be relocated to their original communities; And if there is a problem of "Ascription Lag", that is, a node A is assigned to the original community of node B, while the node B was assigned to another community C, then node A will be reassigned to community C in good time before the community compression of the 2nd stage.

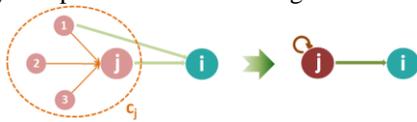

Fig. 4. Parallel computation during community compression

c) Parallelization for the 2nd stage

The compression process can be parallelized directly on the *EdgeTriplet* RDDs. As shown in Fig. 4, if nodes 1、2、3 are assigned to the community where node j originally located, the weights of orange edges are transformed into that of the self-pointed edge in the new graph, and all edges of nodes 1, 2, 3, j connected with another node i are merged into the new edge between i and j. This process can be achieved by picking out the triplets associated with community $c_j$ for node i and j respectively during map phase by using *edges.filter* function, with related edge weights being merged during the reduce phase by using *aggregateMessages* function.

### D. Money laundering risk quantization for Communities

So far the Parallelized version of community division algorithm for ML has been realized. The next thing to do is sorting communities by their ML risk scores. Generally speaking, a community is of higher ML risk if it has an abnormal volume of transfer, more complex transfer structure or more concentrated trading time. Here, the calculation of the temporal risk will be emphatically discussed because the volume and complexity have already been involved during the community detection process.

In information theory, greater entropy indicates a higher uncertainty of information. The value of entropy is only affected by the distribution of variables, regardless of the specific value of the variable itself [17]. Thus a variable called "temporal entropy" is calculated here to measure the transfer time concentration. The average time point $\bar{T}$ of a community is computed at first, and the absolute interval between each transfer time point and $\bar{T}$ is defined as $\Delta T$. Then each transaction can be divided into corresponding segment according the value of $\Delta T$. Finally, the ratio of total transactions in each segment is calculated.

As shown in Fig. 5, the percentage of transactions in the segment of $0 \leq \Delta T < T_1$ is $P_1$, while it is $P_2$ in that of $T_1 \leq \Delta T < T_2$. The rest can be done in the same manner for all segments, and they should obey: $\sum_{i=1}^{n} P_i = 1$. Then the temporal entropy can be defined as: $H_C = -\sum_{i=1}^{n} P_i * \log_2 P_i$. Supposing there is a community k with a total node numbers $V_k$, edge numbers $E_k$, money amount $M_k$, average node degree $\bar{D}_k$ and temporal entropy $H_k$. Then ML risk score for this community can be measured by following formula: $\psi_k = e^{\omega_V \cdot \bar{\bar{V}}_k + \omega_E \cdot \bar{\bar{E}}_k + \omega_M \cdot \bar{\bar{M}}_k + \omega_D \cdot \bar{\bar{D}}_k + \omega_H \cdot \bar{\bar{H}}_k}$, where $\bar{\bar{V}}_k, \bar{\bar{E}}_k, \bar{\bar{M}}_k, \bar{\bar{D}}_k, \bar{\bar{H}}_k$ are the standardization of corresponding variables for community k. Then the communities with relatively higher $\psi_k$ will be paid more attention to, and MCSs with more suspicious communities will be analyzed or reported further.

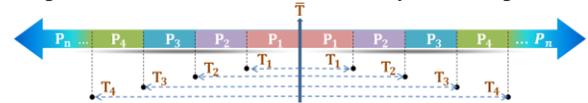

Fig. 5. Transfer period segmentation for calculating temporal entropy

## III. EXPERIMENTS AND RESULTS

The entire process has been verified with the ability to recognize the outliers from majority normal information. The solution was implemented in Spark-1.6.1. All the data were stored on HDFS cluster based on Cloudera CDH-5.9.0. The experimental cluster consists of 100 nodes, where each node contains an Intel Xeon CPU E5-2620 at 2.00GHz CPU and 8 GB RAM. About 10 million of real transfer records were extracted from the transactions in the first week of November 2016 to form a graph. The size of graph was scaled down to about 45% of the original one after filtering isolated edges. The remaining graph was divided into different MCSs after merging edges in 5 min.

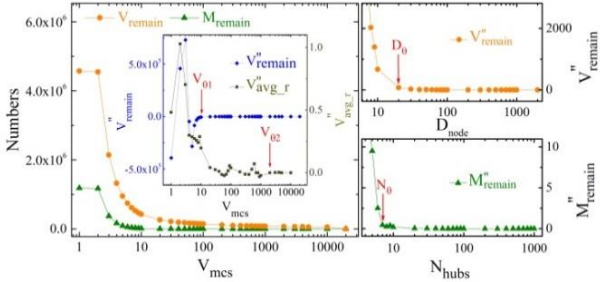

Fig. 6. Determination of threshold parameters from relevant variation trend curves.

As can be seen from the left part of Fig. 6, the orange circle line labeled as $V_{remain}$ represents the number of remaining vertices after being filtered by the threshold scale of MCSs $V_{mcs}$, and the green triangles labeled as $M_{remain}$ represents the number of remaining MCSs after filtered by $V_{mcs}$. The parameter thresholds were determined at the point where each curvature variation speed trend curve (calculated by the second derivative) becomes close to 0. The blue diamond line in inset shows the curvature variation speed trend of $V_{remain}$, labeled as $V''_{remain}$. It vibrates sharply until $V_{mcs}$ exceeds 10. The average scale of the remaining MCSs can be calculated as: $V_{avg\_r} = V_{remain}/M_{remain}$. The brown square line shows the curvature variation speed trend of $V_{avg\_r}$, labeled as $V''_{avg\_r}$. It becomes stable when $V_{mcs}$ passes 2000. Thus, the conditions for filtering the MCSs by scale can be determined as: $10 < V_{mcs} < 2000$. The curvature variation speed trend of remaining vertices *vs.* degree threshold $D_{node}$ has been shown in the upper set of the right part of Fig. 6. The corresponding turning point set threshold

$D_\theta$ to 20 in a similar way. The parameter $N_\theta$ can be determined to 7 from the lower set of the right part of Fig. 6 after the definition of $D_\theta$, according to the variation trend of remaining MCSs number with increasing the hub node scale threshold for a MCS.

After being filtered by all above conditions, a subgraph with 30718 vertices and 135248 edges separated by 74 MCSs was picked out as relative suspicious collection. Community detection was then carried out to further determine the anomaly degree for each MCS. A serial version of original Louvain algorithm based on the primitive edge weight $W_{Bi}$ was first tested. This was done on a single node programed in python, with the subgraph been divided into 736 communities in 38.64s. Then the parallelized version of original Louvain algorithm was implemented. The subgraph was divided into 765 communities in 5.73s with a modularity of 0.385. It achieves a very similar distribution of the serial version in a much shorter time. At last, the parallelized version of TD Louvain algorithm was implemented. It divided the subgraph into 536 communities in 5.97s with a modularity of 0.572, indicating there is no obvious decline in speed despite of the increased complexity and performance.

ML risk score $\psi_k$ for each community k was then calculated and sorted. As shown in Fig. 7, $\psi_k$ are plotted in a descending order, and the first derivative for the risk score $\psi'_k$ is plotted in orange curve. Community percentile curve with risk score are plotted in red circles. Some special percentile points such as 95%, 90% and 80% can be used to determine the suspicious levels. In a general process, communities in the range of 95~100%, 90~95% and 80~90% can be labeled as level 1, 2 and 3. MCSs with multiple communities at corresponding level can be defined as relative suspicious ones according to business requirements. Coincidentally, the suspicious range boundary around 45 inferred from the percentile 90% is very close to the turning point of the derivative curve, after which point the risk score drops regularly at a relatively slow speed. All these phenomena indicate that crucial point 45 can distinguish well whether a community is associated with ML gangs. In addition, the 45 communities with highest risk scores are distributed in 13 MCSs, as shown in the upper part of Fig. 8 with different colors. After further artificial invetigations by internal risk manage department of UnionPay, 9 among the 13 MCSs were reconfirmed suspicious, and a batch of typical nodes with large weights in these MCSs are reported to the China anti-money laundering monitoring and analysis center (CAMLMAC) of the People's Bank of China (PBC).

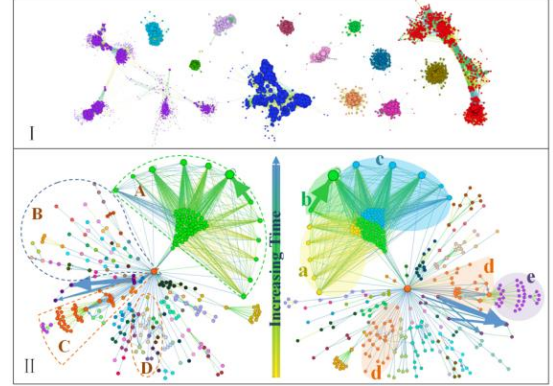

Fig. 8. Ⅰ: Final MCSs with suspicious communities; Ⅱ: Comparison between the results of original weighted Louvain algorithm and TD Louvain algorithm.

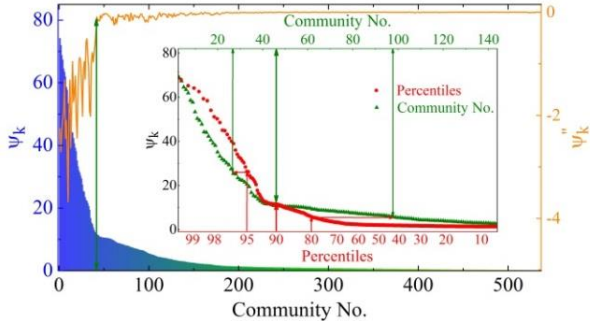

Fig. 7. ML risk level partitioned by risk score distribution

In order to prove the effectiveness of TD Louvain algorithm, one of the suspicious MCSs with 343 vertices and 1050 edges are discussed in detail here as a typical demonstration. The lower left part of Fig. 8 shows the results of the parallelized original Louvain algorithm. Nodes ascribed to different communities are drawn in distinct colors. Moreover, the edges are also drawn in progressive color with increasing time, that is to say, the edge in yellow happens earlier than that in blue. The MCS has been divided into 56 communities. The lower right part of Fig. 8 shows the TD Louvain algorithm result on the same MCS, with the MCS been divided into 31 communities. The distributions appear quite different for the results between TD Louvain algorithm and original one. It can be found that the original community A has been split into communities a, b and c, whose distributions are very in consistent with their transferring time. The original part C and D evolve into the part d and e when using the TD Louvain algorithm. It can be found from edge colors that most of transfers in new part e happens in earlier time, and was included the same group. In addition, it can be clearly seen from the original part that B and D show a quite disorganized distribution, with nodes spreading alternately even on a single branch. The obvious irrationality is well corrected in TD Louvain algorithm by showing a more reasonable division. In a word, TD Louvain algorithm not only make the distribution of fragmented nodes more structured, but also help dividing a large complex network into smaller groups with more explicit meanings.

## IV. CONCLUSIONS

In this paper, we presented a sophisticated solution to find transfer communities with high ML risks in massive transaction networks. Firstly, a whole transaction graph is built by merging edges. The next, transfers with less ML possibility are filtered out by selecting suspicious MCSs. Then a TD Louvain algorithm combined with AML patterns is proposed and implemented on remaining MCSs. The subgraph is further divided into different communities with their ML risk scores calculated. Finally, MCSs containing multi communities at high risk levels are further investigated and reported. Note that all these procedures are implemented on the distributed platform of Spark, and TD Louvain algorithm has also been parallelized and optimized. The results demonstrate that our solution can help to find out criminal gangs with high ML risks in massive transaction networks efficiently and intelligently.

ACKNOWLEDGMENT

This research was supported by National Engineering Laboratory for Electronic Commerce and Electronic Payment, sponsored by High-Tech Service Industry R&D and Industrialization Project of National Development and Reform Commission ([2014] 648 and [2015] 289), Shanghai Sailing Program 17YF1425800 and Pudong New District Science & Technology Development Postdoctoral Fund.